\documentclass[english]{bullsrsl}[2022/06/15]
\usepackage[latin1]{inputenc}
\usepackage[T1]{fontenc}
\usepackage{natbib} 
\usepackage{graphicx}

\RequirePackage{hyperref}

\begin{document}
\title{Gaps in the Main-Sequence of Star Cluster Hertzsprung Russell Diagrams}

\author[affil={}]{Priya}{Hasan}
\affiliation[]{Maulana Azad National Urdu University, Hyderabad, India 500032}
\correspondance{priya.hasan@gmail.com}
\date{31 May 2023}
\maketitle


%

\begin{abstract}
 The presence of gaps or regions of small numbers of stars in the main sequence of the Hertzsprung Russell  Diagram (HRD) of star clusters has been reported in literature. This is interesting and significant as it could be related to  star formation and/or rapid evolution or instabilities.  In this paper, using Gaia DR3 photometry and confirmed membership data, we explore the HRD of nine open clusters  with reported gaps, identify them and assess their importance and spectral types. 
\end{abstract}

\keywords{HRD, star clusters, Gaia DR3, stellar evolution}

\section{Introduction}
The Hertzsprung Russell Diagram (HRD)  of star clusters is the holy grail to understanding stellar evolution and populations. It is a snapshot of stellar lives as a plot of color (temperature) versus magnitude (luminosity). The precise position of a star can be used  to find various parameters of a star including its size, metallicity, and evolutionary state. The HRD traces stars at various phases of evolution along  the main sequence and  as they turn off to the giant branch and beyond. The HRD has been used to find the distances, ages and reddening of star clusters. 

The European Space Agency Gaia mission has provided unprecedented
sub-milliarcsecond parallax precision for over a billion stars \citep{2016, gaiacollaboration2022gaia} that can be utilized to
study the precise locations of individual stars on the HRD as well as populations of stars. The accurate, all-sky data produces an HRD that shows previously unknown features. 

Gaps or regions of low density of stars in the HRD  have been reported by various authors \citep{ 1971Obs....91...78H, 1974ApJ...194..629B,Kjeldsen91,Rachford_2000} and could be important milestones of stellar evolution. 
In this paper, we present a detailed study of main sequence gaps in the HRD of a sample of nine clusters of ages ranging from log $t= 7.09- 9.63$ and at distances $889-2773$ pc using Gaia DR3 data. We use membership data and parameters from \cite{2020A&A...640A...1C}.
We identify the gaps, assess their statistical significance using the $\chi^2$ test and identify their spectral types.

\section{Reported Gaps in Literature}
Main sequence gaps in HRD were reported in literature \citep{Kjeldsen91,1978BASI....6...37S, 1974ApJ...194..629B,Rachford_2000} and are listed in Table \ref{gaps}. A gap was  also found by \cite{Jao_2018} in Gaia DR2 data at $G \approx 10$. The gap is very narrow ($ \approx$ 0.05 mag) and is near the region in the HRD  where M dwarf stars transition from partially to fully convective, near spectral type M3.0V. 
 
\begin{table}[ht]
\centering
\begin{minipage}{88mm}
\caption{Main Sequence Gaps \citep{Kjeldsen91}}
\label{gaps}
\end{minipage}
\bigskip
\begin{tabular}{lllllll}
\hline

\textbf{Structure}& \textbf{$M_V$} &\textbf{$(B-V)_0$} &\textbf{$\Delta M_V$}    &\textbf{$\Delta(B-V)_0$} &\textbf{Sp Type} &\textbf{Temperature (K)} \\
\hline
Mermilloid & 0.0 & -0.12 & 0.25 & & B8V & 12300\\
Canterna Gap & 1.0 & -0.05 & 0.20 & & A1V  & 9330\\
A-bend& 1.3 & -0.02 & 0.7 & 0.05 & A2V & 9040\\
A-group & 1.5 & 0.00 & 0.7 & 0.05 & A3V & 8750\\
M11 gap & 1.7 & 0.05 & 0.5 & 0.5 & A4V & 8480\\
Bohm-Vitense Gap & 2.8 & 0.25 & 0.3 & 0.05 & F0V & 7350\\
NGC 6134-IC4651 gap & 4.5 & 0.5 & 1.0 & 0.15 & G2V & 5800\\
\hline

\hline
\end{tabular}
\end{table}
 
\section{Cluster Sample}
     
We selected a  sample of nine clusters as clusters with confirmed gaps from literature. These are NGC~2169,   
NGC~2360, NGC~1778, NGC~6939, NGC~3680, NGC~2682, Trumpler~1, NGC~2420 and NGC~6134. We  used  the following cluster parameters \cite{2020A&A...640A...1C} shown in Table \ref{clusterdata} to convert magnitudes to absolute scale. The table shows the coordinates of these clusters (RA and Dec), the angular diameter ($r50$) which is the radius that contains half the number of members from the same reference, the logarithm of age $log \ t$, the extinction $A_V$, distance modulus $DM$ and the distance to the cluster in parsecs. 
\begin{table}[ht]
\centering
\begin{minipage}{88mm}
\caption{Cluster  parameters \citep{2020A&A...640A...1C}\\}
\label{clusterdata}
\end{minipage}
\bigskip
\begin{tabular}{llllllll}
\hline
\textbf{Cluster}& \textbf{RA} &\textbf{Dec} &\textbf{Ang.Dia }    &\textbf{log $t$} & \textbf{$A_V$}&\textbf{DM} & \textbf{Distance} \\

&(deg) &(deg) &(deg)    & & (mag)& (mag) & (pc) \\

\hline  
NGC 2169 &	92.13  & 13.95 & 0.076 & 7.09 & 0.85 & 10.15 & 1072\\  
NGC 2360 &  109.44 &-15.63 & 0.154 & 9.01 & 0.39 & 10.25 & 1122 \\ 
NGC 1778 & 77.03 & 37.02 & 0.112 & 8.25 & 0.87 & 11.11 & 1663 \\
NGC 6939 &  307.9  & 60.65 & 0.123 & 9.23 & 0.85 & 11.3  & 1815  \\  
NGC 3680 & 171.39  &-43.24 & 0.149 & 9.34 & 0.1  & 10.15 & 1072 \\  
NGC 2682 & 132.85  & 11.81 & 0.167 & 9.63 & 0.07 & 9.75  & 889  \\
Trumpler 1  & 23.92   & 61.28 & 0.031 & 7.46 & 1.63 & 12.22 & 2773 \\ 
NGC 2420 & 114.6   & 21.58 & 0.053 & 9.24 & 0.04 & 12.06 & 2587  \\
NGC 6134 & 246.95  & -49.16& 0.156 & 8.99 & 0.87 & 10.36 & 1182  \\  

\hline
\end{tabular}
\end{table}

\section{Analysis}

We use membership data from \cite{2020A&A...640A...1C} for our sample of nine clusters. 
As described in \cite{donada2023multiplicity}, we find the absolute magnitude and color:
$$ G= G -\mu -0.89 *A_V$$  
$$(BP-RP)_0= (BP-RP) -0.89/1.85*A_V$$ .
We plot the color magnitude diagrams (Fig. \ref{cmd})  and the luminosity functions (Fig. \ref{lf}) to identify possible gaps in the HRD listed in Table \ref{gapslist}. 

\bigskip
\begin{figure}
\centering
\includegraphics[width=0.8\textwidth]{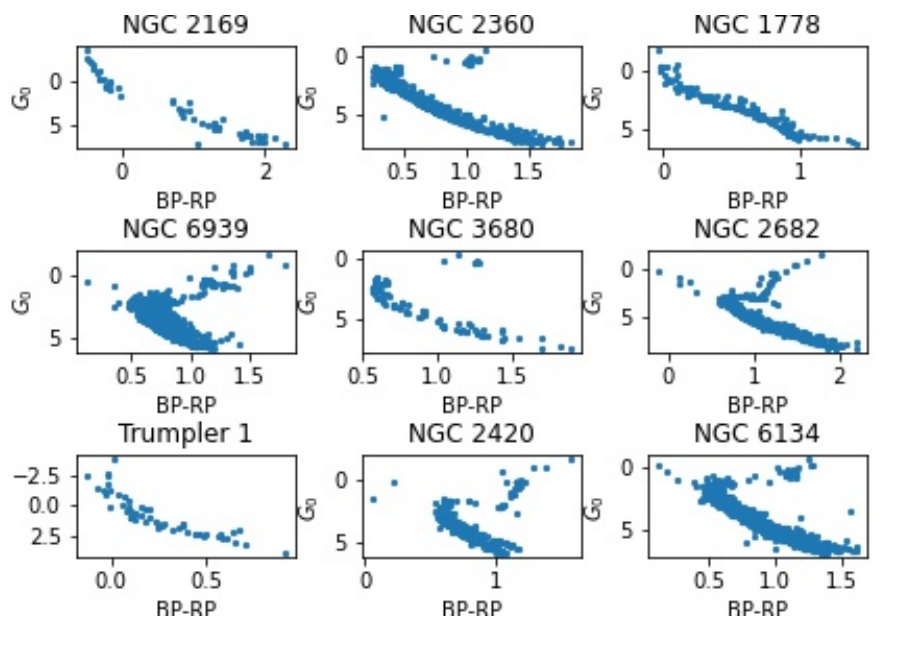}
\bigskip
\begin{minipage}{12cm}
\caption{HRD of the sample of 9 clusters.}
\end{minipage}
\end{figure}

\begin{figure}
\centering
\includegraphics[width=0.8\textwidth]{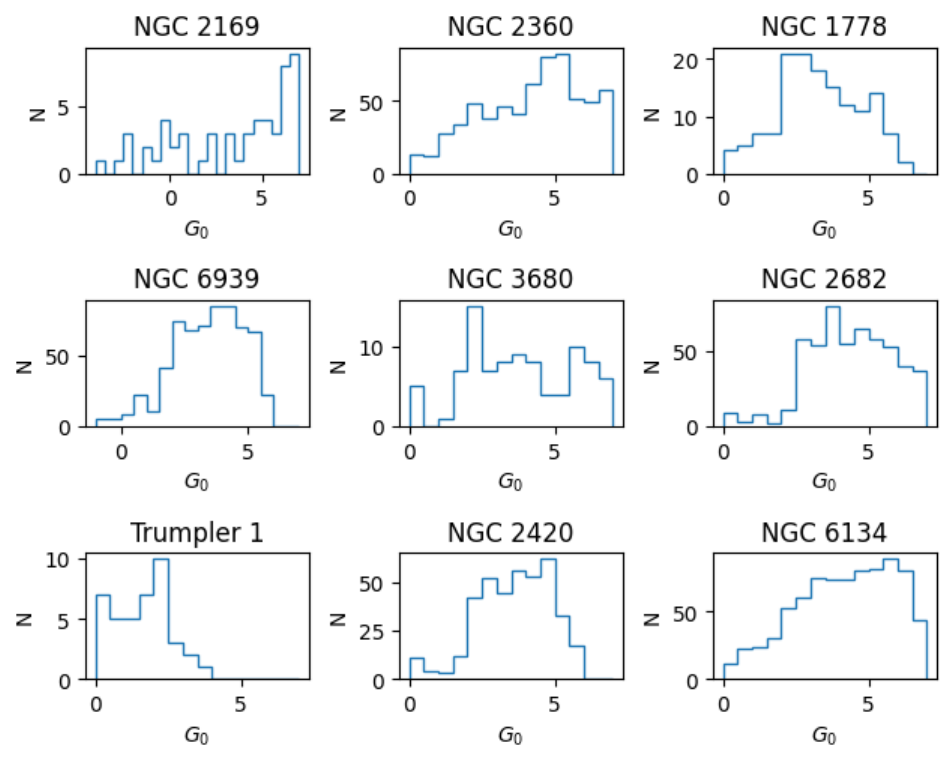}
\bigskip
\begin{minipage}{12cm}
\caption{Luminosity functions of the sample of 9 clusters.}
\end{minipage}
\end{figure}

The likelihood that the observed gap represents a chance variation can be estimated as follows. 
For the identified gaps, we calculate $\chi^2=\frac{(N-N_o)^2}{N}$, where $N$ is the expected number of stars and $N_0$ is the observed number of stars as described in \cite{1971Obs....91...78H}. We find the expected number as the average of the numbers before and after the gap.  $\chi^2$  is related to $p$ that is the probability that the gap is an chance event. For a $\chi^2$=4.0, with one degree of freedom, the $p$ value is 0.05. This means that the probability of the gap being significant is 1-0.05 = 0.95, that is  95 \%. This implies that a smaller value implies a higher chance of the gap being significant. Table \ref{gapslist} lists the gaps found in our sample with their spectral Types and significance. We notice  the gaps we found  are of similar spectral types as described in Table~\ref{gaps}.

\begin{table}
\centering
\begin{minipage}{88mm}
\caption{Details of Gaps found}
\label{gapslist}
\end{minipage}
\bigskip
\begin{tabular}{lllllllll}
\hline
\textbf{Cluster}& \textbf{$G_{median}$} &\textbf{$G_{BP-RP} mean$} &\textbf{$N$}    &\textbf{$N_0$} &\textbf{$\chi^2$}&\textbf{p}  & \textbf{T(K)} & \textbf{Sp Type} \\
\hline 
NGC 2169 &	2.2 & 0.3 &4	 &0	& 4.0 &0.05 &6852 & F8V\\
         &  5.6 & 1.45&7.5   &3 & 2.7 & 0.1 &3631 & K1V \\
 \hline  
NGC~2360 & 0.8 & 0.12 & 17.4 & 12   & 1.67 &0.196 & 8550 & A3V \\ 
         & 1.7 & 0.3  & 21.6 & 17.8 & 0.67 & 0.41  &7500& A8V \\ 
         & 2.8 & 0.45 & 24.3 & 18.5 & 1.38 & 0.24 &7030& F1V \\ 
 \hline 
NGC 1778 & -1.11 & -0.03 & 3 & 0 & 3 & 0.08 &9700 & A0V\\
 & 1.5 & 0.14 & 14 & 7 & 7 & 0.008 & 8550 & A3V\\ \hline
NGC~6939 &3.0 & 0.5 & 80 & 70 & 1.25 & 0.26 &6720&F3V \\  
      \hline 
NGC~3680  & 5  & 0.95  & 8.5 & 4 & 2.38& 0.1229 & 5280  & K0V\\
\hline 
NGC~2682& 3.07 &0.7 &67.5&  54.5 &2.5 & 0.1138 & 6040 & F9V   \\
        & 4.2 &  0.77 & 80 & 55.6    & 7.44   & 0.0064 & 5880 & G1V \\ \hline 
Trumpler~1 & -2.0 & -0.08 & 2.5 & 1 & 0.9 & 0.34 & 10400 & B9.5V  \\ 
&-0.25 & 0.018 & 5 & 2 & 1.8 & 0.18& 9200 & A1V\\
&1 & 0.121 & 6 & 5 & 0.16& 0.69& 8550 & A3V\\
&1.75 & 0.25 & 4.7 & 3 & 0.617&0.43& 7800 & A7V\\ \hline 
NGC 2420 & 3.25 & 0.55 & 54& 44 & 1.85& 0.17 & 6640 & F4V \\
         & 4.25 & 0.8 & 59 & 53 &0.61 & 0.43 & 5770 & G2V \\ \hline 
NGC~6134 & 2.5 & 0.5 & 30.6 & 27 & 0.42 &0.62& 6720 & F2V\\
&3.15 & 0.6     & 37.4     & 31   & 1.09 & 0.29&6400 & F5V \\
& 3.6 & 0.675 & 39&34.4 & 1.09 & 0.29& 6150 & F8V\\ 
&4.09 & 0.748 & 39.5 & 33.5 & 0.91 & 0.34& 5920 & G0V\\
& 4.62 & 0.83 & 44.6 & 28 & 6.17 & 0.012&5660 & G5V\\ \hline         
\hline
\end{tabular}
\end{table}

\section{Conclusions}
 In this paper, we use Gaia DR3 data and membership data  of \cite{2020A&A...640A...1C} to study gaps in the main sequence of the HRD of star clusters. We use the $\chi^2$ test to find the significance of the gaps. We compare the spectral types of earlier detections and find that they agree with our present results. 
Gaps were reported by \cite{Jao_2018} in Gaia DR2 data for M dwarfs. In our sample, the membership data used is available only till apparent $G$ magnitude 18 and does not include M dwarfs, we go to a spectral type of upto G, therefore we don't find that in our data.  
A more detailed study of HRD of star clusters is necessary to characterise these gaps and study them in more detail.

\begin{acknowledgments}
This work has made use of data from the European Space Agency (ESA) mission
{\it Gaia} (\url{https://www.cosmos.esa.int/gaia}), processed by the {\it Gaia}
Data Processing and Analysis Consortium (DPAC,
\url{https://www.cosmos.esa.int/web/gaia/dpac/consortium}). Funding for the DPAC
has been provided by national institutions, in particular the institutions
participating in the {\it Gaia} Multilateral Agreement.
\end{acknowledgments}

\begin{furtherinformation}

\begin{orcids}
\orcid{0000-0002-8156-6940}{Priya}{Hasan}


\end{orcids}


\begin{conflictsofinterest}
The authors declare no conflict of interest.
\end{conflictsofinterest}

\end{furtherinformation}

\bibliographystyle{bullsrsl-en}

\bibliography{S04-P12_HasanP}

\end{document}